\documentstyle[prl, aps, amssymb]{revtex}
\begin{document}
\draft
\wideabs{
\title{$^{63}$Cu NQR Evidence for Spatial Variation of Hole 
Concentration in La$_{2-x}$Sr$_{x}$CuO$_{4}$}
\author{P.M. Singer, A.W. Hunt and T. Imai}
\address{Department of Physics and Center for Materials Science and 
Engineering, M.I.T., Cambridge, MA 02139}
\date{\today}
\maketitle

\begin{abstract}
We report experimental evidence for the spatial variation of hole 
concentration $x_{hole}$ in the high $T_{c}$
superconductor La$_{2-x}$Sr$_{x}$CuO$_{4}$ ($0.04\leq x \leq 0.16$) by
using $^{63}$Cu NQR for $^{63}$Cu isotope 
enriched samples. 
We demonstrate that the extent of the spatial variation  of the local hole concentration 
$\bigtriangleup x_{hole}$ is reflected 
on $^{63}1/T_{1}$ and deduce the temperature dependence. 
$\bigtriangleup x_{hole}$ increases below $500 - 600$K, and reaches 
values as large as $\bigtriangleup x_{hole}/x \sim 0.5$ below $\sim 150$K. 
We estimate the
length scale of the spatial variation in $x_{hole}$ to be $R_{hole} \gtrsim 
$ 3 nm from analysis of the NQR spectrum.

\end{abstract}
}
The mechanism of high $T_{c}$ superconductivity remains enigmatic 
even after 15 years of its discovery\cite{BM}. It is well known 
that the central idea in the search of high $T_{c}$ materials was 
to look for materials with polaronic effects caused by 
coupling between electrons and the lattice.  However, most recent theoretical 
debates are based on the assumption that the
holes are homogeneously doped into CuO$_{2}$ planes, even 
if hole-doping is achieved by the substitution of ions with 
different ionicity which in effect creates alloys. A clear and widely accepted counter example 
against such a simplified picture is 
the stripe phase in Nd co-doped La$_{2-x}$Sr$_{x}$CuO$_{4}$, where the 
Coulomb potential from the distorted lattice slows down spin 
and charge density waves \cite{Tranquada}.  
It is also known that the high mobility of the oxygen atoms results in
electronic phase separation \cite{Sokol,kivelson,castro} between the superconducting and 
antiferromagnetic phase in super-oxygenated 
La$_{2}$CuO$_{4+\delta}$ as evidenced by $^{139}$La NMR 
(nuclear magnetic resonance) \cite{stat}. A natural question to ask 
is whether similar situations arise in other materials albeit in a 
less robust manner \cite{burgy}. In fact, there is ample evidence for some sort 
of spatial inhomogeniety in the CuO$_{2}$ plane from earlier 
NMR and NQR (nuclear quadrupole resonance) studies in La$_{2-x}$Sr$_{x}$CuO$_{4}$ 
\cite{yoshimura,tou,fujiyama,hunt,singer,curro,haase,julien} such as splitting of the 
$^{63}$Cu NQR and NMR lines due to ineqivalent Cu sites with 
different EFG (electric field gradient) tensors 
\cite{yoshimura,stat}, drastic broadening of the $^{63}$Cu NQR line at 
low temperatures \cite{tou,hunt}, and $^{63}$Cu NMR line broadening due 
to the modulation of orbital shifts \cite{haase}.
More recent studies using $^{63}$Cu NQR and NMR wipeout 
\cite{hunt,curro,julien} have characterised the glassy nature of the 
slowing down of the stripe inhomogeneity below temperatures $\sim$100K. 
However, no clear picture has emerged that discerns and relates the 
effects of genuine electronic phase separation, stripe modulation, and 
random substitution of donor ions. 
On the other hand, a recent STM (scanning tunnelling 
microscopy) study on the surface state 
of Bi$_{2}$Sr$_{2}$CaCu$_{2}$O$_{8+\delta}$ cleaved at low 
temperatures revealed nano-scale spatial variations of the electronic 
state\cite{pan}. Whether such nano-scale modulations are universally 
observable in the bulk and other high $T_{c}$ materials remains 
to be seen, but the STM results have enhanced the interest in the potential impact of 
the spatial inhomogeniety of the electronic properties in the CuO$_{2}$ plane.

In this Letter, we utilize a new simple trick based on the local nature 
of the $^{63}$Cu NQR technique to probe the spatial 
inhomogeniety of the CuO$_{2}$ plane. Unlike scattering techniques,
NQR does not require spatial coherence over tens of nm's, hence it is an ideal 
technique to probe spatial variations in the local hole concentration 
$x_{hole}$ at short length scales. 
Our main results are presented in Fig. 1, where we plot the 
temperature dependence of the extent of the spatial variation
$ \bigtriangleup x_{hole}$ as measured by the distribution
in $^{63}1/T_{1}$ (data with solid line) and by
analysis of the NQR spectrum (dashed lines). Contrary to common assumptions that doped 
holes are uniformily distributed in the CuO$_{2}$ planes (i.e. 
$\bigtriangleup x_{hole}$=0), we find that a 
non-zero value of $\bigtriangleup x_{hole}$ exists at all temperatures 
and furthermore shows a factor 2-3 increase 
from 500-600K down to $\sim$150K.
We also demonstrate that the NQR spectrum is consistent with a model where the 
spatial variation in $x_{hole}$ takes the form of patches in the 
CuO$_{2}$ plane, where some patches are more metallic and some more 
insulating, with a patch radius $R_{hole} \gtrsim $3 nm.

We shall now describe the details of the proceedure used to obtain $\bigtriangleup 
x_{hole}$ shown in Fig. 1.  
The nuclear spin-lattice relaxation rate $^{63}1/T_{1}$ is given by the formula:
\begin{equation}
	 ^{63} \frac{1}{T_{1}}  = \frac{2}{g^{2} \mu_{B}^{2} \hbar} 
	\sum_{{\bf q}} |^{63}A({\bf q})|^{2} S({\bf q},\omega_{n})	
\label{T1}
\end{equation}
where $\omega_{n}$ is the NQR frequency,
$^{63}A({\bf q})$ is the hyperfine form factor, and $S({\bf q},\omega_{n})$ is the dynamic
stucture factor of the Cu electron spins \cite{moriya}.
$^{63}1/T_{1}$ measures the $local$ low frequency
Cu spin fluctuations at $\omega_{n}$.
Because of the experimental ease, the temperature dependence of 
$^{63}1/T_{1}$ is generally measured at the peak of the spectrum. 
However, in various materials including 
La$_{2-x}$Sr$_{x}$CuO$_{4}$ \cite{fujiyama}, 
$1/T_{1}$ often depends on frequency across the resonance spectrum due 
to a spatial distribution of the electronic states. 
A typical example is shown in Fig. 2a for $x=0.115$. For convenience we have 
defined $1/T_{1}^{0}$, $1/T_{1}^{+}$, $1/T_{1}^{-}$ and $1/T_{1}^{-1/10}$ 
as $^{63}1/T_{1}$ measured at the peak, upper half intensity, lower 
half intensity and lower one tenth intensity of the A-line respectively.

The temperature dependence of $1/T_{1}^{+}$, $1/T_{1}^{-}$ and 
$1/T_{1}^{-1/10}$ is plotted in Fig. 3. All data were taken
in the temperature region above Cu wipeout \cite{hunt}, where full NQR 
signal intensity is observable. For comparison, Fig. 3 also 
shows $1/T_{1}^{0}$ for a variety of samples which all show 
agreement with previous work \cite{imai}.  We note that all of our results 
are insensitive to RF pulse width.
The most surprising discovery of the present work is that $1/T_{1}^{+}$, $1/T_{1}^{-}$ and 
$1/T_{1}^{-1/10}$ show {\it qualitatively different} temperature 
dependence. For example, in the $x=0.07$ sample, $^{63}1/T_{1}$ 
measured at the half intensity of the lower (upper) frequency side of 
the NQR spectrum exhibits semiquantitatively the same behaviour as 
$1/T_{1}^{0}$ for $x=0.04$ ($x=0.115$). This is consistent with 
the fact that the lower (upper) frequency side of the 
NQR spectrum for
$x=0.07$ roughly coincides with the peak NQR frequency of $x=0.04$
($x=0.115$) as shown in Fig. 2b. Our finding immediately establishes that 
within a single sample with a fixed nominal hole concentration $x$,
there are some segments with higher and lower hole 
concentrations $x_{hole}$.

By comparing $1/T_{1}^{\pm}$ with smoothly interpolated values of 
$1/T_{1}^{0}$ for various samples, we can estimate the deviation 
$\bigtriangleup x_{hole}$ of local hole concentration from the spatial 
average, $x$. 
For example, $1/T_{1}^{-}$ for $x=0.07$ at 
200K 
is close to $1/T_{1}^{0}$ for $x \sim 0.045$, implying that the lower frequency 
side of the $x=0.07$ spectrum
corresponds to resonance from segments with $\bigtriangleup x_{hole} \sim - 
0.025$. 
Similarly, $1/T_{1}^{+}$ for $x=0.07$ is close to $1/T_{1}^{0}$ for 
$x \sim 0.10$, implying the upper frequency side corresponds to 
segments with $\bigtriangleup 
x_{hole} \sim + 0.03$. Using $1/T_{1}^{-1/10}$ instead
results in an overall $30 - 40 \%$ 
increase in $\bigtriangleup x_{hole}$ and does not effect our 
conclusions. Another important feature in Fig. 3 is that 
$1/T_{1}^{\pm}$ cuts through the lines of $1/T_{1}^{0}$
with decreasing temperature. This implies that $\bigtriangleup x_{hole}$ 
increases with decreasing temperature, as shown in Fig. 1. Since 
the magnitude of $\bigtriangleup x_{hole}$ estimated from $1/T_{1}^{+}$ 
and $1/T_{1}^{-}$ is identical within experimental uncertainties, we 
plot the magnitude of $\bigtriangleup x_{hole}$.
To the best of our knowledge, our $^{63}$Cu NQR data in Fig. 1 is the 
first of its kind to detect the temperature dependence of the 
inhomogeneous distribution of the hole concentration in La$_{2-x}$Sr$_{x}$CuO$_{4}$
or any other high $T_{c}$ materials with quenched disorder.

One important observation about Fig. 2a is that the B-line 
\cite{yoshimura} shows a very 
similar distribution in $^{63}1/T_{1}$ to the A-line. 
The B-line itself originates from structural effects 
\cite{pleb}. More specifically, a B-site corresponds to a Cu nucleus with 
a Sr$^{2+}$ ion
situated either directly above or below it, as supported in Fig. 2c where 
the fractional intensity of the B-line is shown to increase as $x$. 
We recall that the $^{63}$Cu NQR spectrum in 
La$_{2-x-y}$Eu$_{y}$Sr$_{x}$CuO$_{4}$ for $^{63}$Cu isotope enriched 
samples \cite{hunt} shows a third structural peak (the C-line) whose 
fractional intensity is equal to $y$. The C-line corresponds to a Cu 
nucleus directly above or below a Eu$^{3+}$ ion. 

The similarity of the B-line to the A-line distribution of $^{63}1/T_{1}$
rules out the possibilty that the distribution in the hole concentration 
$\bigtriangleup x_{hole}$ is concentric about a B-site.
Rather, it indicates that the spatial modulation of $\bigtriangleup 
x_{hole}$ consists of patches that cover both A and B-sites equally and
the size of each patch is larger than the average 
Sr$^{2+}$ to Sr$^{2+}$ distance of $\sim$1(2) nm for $x=0.16(0.04)$. 

We emphasize that the distribution in 
$^{63}1/T_{1}$ is {\it not} caused by flaws
in our single phased ceramic samples. $^{63}$Cu NMR and NQR in high quality 
single crystals of both La$_{2-x}$Sr$_{x}$CuO$_{4}$ \cite{provided}
and La$_{2}$CuO$_{4+\delta}$ \cite{ylee} exhibit 
similar results. EMPA (electron micro-probe analysis) on single crystal 
La$_{1.85}$Sr$_{0.15}$CuO$_{4}$ showed that the spatial 
variation of the Sr$^{2+}$ content averaged over the focus area of the 
electron beam of $\sim 1 \mu$m is $\bigtriangleup x_{Sr} \sim 0.01 \ll 
\bigtriangleup x_{hole}$. This rules out the possibility that $\mu$m scale 
inhomogeneities in the Sr$^{2+}$ concentration is the cause of $\bigtriangleup 
x_{hole}$. 
Moreover, earlier NQR measurements by Fujiyama et. al. 
\cite{fujiyama} in La$_{2-x}$Sr$_{x}$CuO$_{4}$ $without$ isotope
enrichment agree with ours. 
We also note that the planar $^{17}$O NMR Knight shift measurements provide 
additional evidence for a spatial distribution in the hole 
concentration. Our systematic measurments for single crystals 
\cite{provided} with $x=0.025$, 0.035, 0.05, 0.115 and 
0.15 showed an overlap of the planar $^{17}$O NMR central transition very similar to 
Fig. 2b.  This means that the local static spin susceptibility has a similar 
spatial variation.

We now turn our attention to the analysis of the $^{63}$Cu NQR 
spectrum. Experimental resolution limited the $\bigtriangleup x_{hole}$ 
data deduced from the distribution in $^{63}1/T_{1}$ above $\sim 500$K
since all $1/T_{1}^{0}$ curves start to merge above this temperature 
(Fig. 2). 
Not only does the spectral analysis enable
us to extend $\bigtriangleup x_{hole}$
to higher temperatures, but we also 
get crucial insight into the length scale of the spatial variation $\bigtriangleup 
x_{hole}$. 

The resonance frequency in NQR is proportional to
the EFG surrounding the Cu nucleus \cite{slichter}. The broad NQR linewidth shown in Fig. 
2a and 2b implies that the local charge environment has a broad 
distribution in the EFG. Possible causes of the broad 
distribution include: the random substitution of La$^{3+}$ with 
Sr$^{2+}$, the spatial variation $\bigtriangleup x_{hole}$, and the lowering of 
the local symmetry by lattice distortions.
In order to simulate these effects, we carried out a point charge 
lattice summation to determine 
the EFG at the Cu nuclear site. The conventional way of incorporating the Sr$^{2+}$ substitution into 
an EFG point charge calculation is to reduce the effective charge of 
the La$^{3+}$ ion to La$^{(3-x/2)+}$ \cite{shimizu}. However, in order to estimate the effect
of the random replacement of La$^{3+}$ with 
Sr$^{2+}$, we ran a simulation of a realistic lattice with 
randomly distributed Sr$^{2+}$ ions in a La$^{3+}$ 
matrix\cite{shielding}. We found that 
the linewidth thus deduced is too narrow to account for the 
experimental result, as shown by the dashed line in Fig. 2a.

Next, we ran the simulation with $\bigtriangleup x_{hole} \neq 0$ using the following procedure: 
we first define an in-plane
circle of radius $R_{hole}$ surrounding the Cu nucleus $\alpha$. The circle 
defines a "patch" of diameter $2R_{hole}$ within which the local hole concentration is determined 
by counting the number $N$ of Sr$^{2+}$ ions  
within the patch. 
The $N$ donated holes from the $N$ Sr$^{2+}$ ions are then uniformly distributed among the 
planar oxygen sites within the patch. The EFG contributions from all 
the point charges surrounding the Cu site $\alpha$ 
is then summed up, the corresponding NQR frequency deduced, and a 
count is placed in the appropriate frequency bin. The lattice is then 
re-randomized, and the whole proceedure is repeated
$\sim 10^{4}$ times until a sufficient number of counts is present to 
reproduce the entire NQR spectrum, including the B-line.
The aforementioned case of the simulation with $\bigtriangleup x_{hole}$=0 
(i.e. homogeneous distribution of holes) 
corresponds to $R_{hole} \sim \infty$ and yields a linewidth much narrower than 
the experimental data. Using $R_{hole} \sim $1.5 nm yields the dashed-dot spectrum 
in Fig. 2a, which is clearly too broad.
Using $R_{hole} \sim $3 nm 
however, results in the best fit shown by the solid lines in Fig. 
2a and 2b. Once the best fit $R_{hole}$ is determined, $\bigtriangleup 
x_{hole}$ is directly obtained from the computation and is presented 
as the dashed lines in Fig. 1.
According to the simulation, the $\sim 25$ \% increase in linewidth from
600K (Fig. 2b) to 300K (Fig. 2a) for $x=0.115$ can be accounted for by
a $\sim 25$ \% decrease in $R_{hole}$ which corresponds to $\sim 45$ \% increase in $\bigtriangleup 
x_{hole}$ from 600K down to 300K. We discuss implications of the temperature 
dependence later. 

We caution that we have ignored potential local tilting of the CuO$_{6}$ 
octahedra caused by subtitution of the smaller 
Sr$^{2+}$ ions, or potential precursor effects \cite{haskel,bozin,braden} 
to the long range LTO (low temperature orthorhombic) transition. 
 Since local octahedron tilting increases the NQR resonance 
frequency \cite{tilt,singer}, potential local distortions of the lattice may 
provide the additional mechanism to broaden the spectrum.
We have ignored 
this and attributed the extra line broadening to $\bigtriangleup x_{hole}$, 
hence $\bigtriangleup x_{hole}$ thus deduced may be an upper bound and  
$R_{hole} \sim 3$ nm maybe a lower bound.
We note that the minimum patch size of $2R_{hole} \geq $6 nm is still larger than the average 
Sr$^{2+}$ ion separation of 1(2) nm for $x = 0.16(0.04)$, which further rules out 
the possibilty that the distribution of the local hole concentration 
is concentric about {\it each} Sr$^{2+}$ ion.
The validity of our simulation breaks down altogether in 
the temperature region where local octahedron tilting 
actually becomes apparent due to long range LTO transition,
as indicated by an increase in $^{63}\nu _{Q}$ \cite{imai,singer}.
The analysis shown in Fig. 1 is thus terminated below this temperature.

Our lineshape analysis sets only the lower bound of the length scale $2R_{hole} \geq $6 nm
of the spatial variation $\bigtriangleup x_{hole}=0.03\sim 0.06$.  On 
the other hand, our aformentioned EMPA analysis 
indicates that the deviation in 
Sr$^{2+}$ concentration is as small as $\bigtriangleup x_{Sr} \sim 
0.01$ averaged over the length scale of $\sim 1 \mu$m.  Furthermore, 
earlier high resolution X-ray 
diffraction studies \cite{takagidelta} showed that $\bigtriangleup x_{Sr} \sim 
0.01$ averaged over the length scale of $\gtrsim 10$ nm.  These 
results imply that the local hole concentration $x_{hole}$ modulates as 
much as $\bigtriangleup x_{hole}=0.03\sim0.06$ within the 
short distance scales of 6-10 nm, but the spatial average of $x_{hole}$ over greater 
length scales is as little as $\sim 0.01$. 
One obvious possibility is the clustering of Sr$^{2+}$ ions with very 
short length scale of 6-10nm.
However the absence of any substantial diffuse 
scattering in neutron diffraction experiments \cite{braden} makes this 
scenario unlikely.  Moreover, the presence of the temperature dependence of 
$\bigtriangleup x_{hole}$ suggests that the temperature
independent quenched disorder caused by the clustering of the Sr$^{2+}$ ions alone cannot entirely 
account for our results. We recall that our 
lineshape analysis is based on a completely random distribution of 
Sr$^{2+}$ ions. We also note that the spin-spin correlation length 
$\xi$ cannot be identified as $R_{hole}$ because $\xi$ is much shorter 
than $2R_{hole}$
and {\it increases} with decreasing temperature \cite{yamada}. 
The latter qualitatively contradicts 
with the {\it decrease} of $R_{hole}$ with decreasing temperature.    
These considerations inevitably lead us to conclude that there 
must exist an electronic mechanism at short length scale which causes the holes to segragate.

One possible mechanism of establishing such a short electronic length scale 
is phase separation (which may or may not have a stripe-like 
structure).  In fact, Ino et. al. \cite{ino} 
reported evidence for phase separation at $\sim 80$K in 
La$_{2-x}$Sr$_{x}$CuO$_{4}$ for $x\leq$0.15 based on photoemission 
experiments, {\it i.e.} the chemical 
potential does not vary with increasing hole doping up to $x=0.15$. In 
this concentration regime, the presence of a stripe instability has been 
firmly established at low temperatures \cite{Tranquada,yamada,hunt}. Our new observation in Fig.1 
{\it may} be a precursor to such a phenomena. However, we observed 
comparable or larger values of $\bigtriangleup x_{hole}$ for
$x=0.20$ and 0.25 where photoemission 
measurements\cite{ino} do {\it not} detect signatures of phase 
separation. Thus an additional mechanism(s) for segregation of holes is 
necessary to account for our results at least above $x\sim 0.15$.      
Interestingly, the temperature dependence of $\bigtriangleup 
x_{hole}$ and the magnitude of $R_{hole}$ may be qualitatively understood based on a toy model of a thermal 
activation type process for the segragation of holes.
Using the in-plane static dielectric constant of $\epsilon_{s} \sim 
30$ and the effective mass $m^{*} \sim 2$ (in units of $m_{e}$) deduced from lightly doped 
La$_{2}$CuO$_{4+\delta}$ \cite{chen}, one obtains a factor $ 
m^{*}/\epsilon_{s}^{2} \sim 1/450$ smaller binding energy 
of E$_{b} \sim 30$meV or E$_{b} \sim 350$K and a factor
$ \epsilon_{s}/ m^{*} \sim 15$ larger Bohr radius of a$_{0} \sim 
1$nm. These temperature and length scales are indeed comparable to our observation. Even 
though this oversimplified model does not include the effects of 
screening for instance, it might suggest that the donated holes 
are bound by a Coulomb type potential created by the Sr$^{2+}$ ions 
rather than
homogeneously doped into the CuO$_{2}$ planes. Needless to say, our toy model analysis employed here above 
$\sim$100 K does not necessarily support or rule out the presence of intrinsic phase separation at 
lower temperatures. 

The work was supported by NSF DMR 98-08941 and 99-71264.


%
%

\begin{figure}
\caption{Temperature dependence of the distribution in local hole concentration 
$ \bigtriangleup x_{hole}$ as deduced by $^{63}$Cu NQR 
$1/T_{1}^{\pm}$ in La$_{2-x}$Sr$_{x}$CuO$_{4}$ ($\blacktriangle$ 
$x=0.04$; $\circ$ $x=0.07$; $\bullet$ $x=0.115$; $\vartriangle$ 
$x=0.16$). Single points at 600K with dotted lines 
are upper bounds deduced from the best fit to the $^{63}$Cu NQR spectra.}
\label{Delta x}
\end{figure}

\begin{figure}
\caption{(a) Frequency dependence of $^{63}1/T_{1}$ ($\square$) across $^{63}$Cu NQR 
spectrum ($\bullet$) in 
La$_{1.885}$Sr$_{0.115}$CuO$_{4}$ at 300K. Dashed, solid and 
dashed-dot spectra are theoretical fits with patch sizes $R_{hole} \sim \infty$, 
3nm and 1.5nm respectively. Dotted lines through $^{63}1/T_{1}$
are a guide for the eye. (b) $^{63}$Cu NQR 
spectra in $x=0.04$ ($\blacktriangle$), 
$x=0.07$ ($\circ$), $x=0.115$ ($\bullet$), and $x=0.16$ ($\vartriangle$) at 600K.  
Solid lines are theoretical fits with $R_{hole}\sim 3$nm. 
(c) Fraction f$_{B}$ of B-line intensity to total intensity (with $T_{2}$ 
corrections) at ($+$) 600K 
and ($\times$) 300K as a function of $x$.}
\label{600K}
\end{figure}

\begin{figure}
\caption{Temperature dependence of $1/T_{1}^{-1/10}$ ($\times$), 
$1/T_{1}^{-1}$ ($\blacktriangledown$), $1/T_{1}^{0}$ ($\circ$), and
$1/T_{1}^{+}$ ($\blacktriangle$) in La$_{2-x}$Sr$_{x}$CuO$_{4}$ where 
$x$ is given in each panel. Solid lines are a guide for the eye, 
and dashed lines are $1/T_{1}^{0}$ for $x=0.00$, 0,02, 0.04, 0.07 
,0.09, 0.115, and 0.16 where $1/T_{1}^{0}$ monotonically decreases 
with increasing $x$.}
\label{4}
\end{figure}


\begin{references}
\bibitem{BM}J.G. Bednorz and K.A. M$\ddot{u}$ller, Z. Physik B {\bf 64}, 189 (1986).
\bibitem{Tranquada}J.M. Tranquada et. al., Nature {\bf 375}, 561 (1995).
\bibitem{Sokol}L.P. Gorkov et al., JETP Letters {\bf 46}, 420 (1987).
\bibitem{kivelson}V.J. Emery et al., Physica C {\bf 209}, 597 (1993).
\bibitem{castro}A.H Castro Neto, Phys. Rev. B {\bf 51}, 3254 (1995).
\bibitem{stat}B.W. Statt et al., Phys. Rev. B {\bf 52}, 15575 (1995).
\bibitem{burgy}J. Burgy et. al., cond-mat/0107300.
\bibitem{yoshimura}K. Yoshimura et al., J. Phys. Soc. Jpn. {\bf 58}, 3057 (1989).
\bibitem{tou}H. Tou et. al., J. Phys. Soc. Jpn. {\bf 61}, 1477 (1992).
\bibitem{fujiyama} S. Fujiyama et al., J. Phys. Soc. Jpn. {\bf 66}, 2864 (1997).
\bibitem{hunt}A.W. Hunt et al., Phys. Rev. Lett. {\bf 82}, 4300 
(1999) and A.W. Hunt et al., Phys. Rev. B {\bf 64}, 134525 (2001).
\bibitem{singer}P.M. Singer et al., Phys. Rev. B {\bf 60}, 15345 (1999). 
\bibitem{curro}N.J. Curro et. al., Phys. Rev. Lett. {\bf 85}, 642 (2000).
\bibitem{haase}J. Haase et. al., Physica C {\bf 341}, 1727 (2000). 
\bibitem{julien}M.-H. Julien et al., Phys. Rev. B {\bf 63}, 144508 (2001).
\bibitem{pan}S.H. Pan et al., cond-mat/0107347.
\bibitem{moriya}T. Moriya, J. Phys. Soc. Jpn. {\bf 18}, 516 (1963).
\bibitem{imai}T. Imai et al., Phys. Rev. Lett. {\bf 70}, 1002 (1993).
\bibitem{pleb}S. Plibersek et al., Europhys. Lett. {\bf 50}, 789 (2000).
\bibitem{provided}Provided by F.C. Chou (M.I.T.); K. Hirota (Tohoku 
University); M. Takaba, T. Kakeshita, H. Eisaki, and S. Uchida (University of Tokyo).
\bibitem{ylee}Y.S. Lee et al., Phys. Rev. B {\bf 60}, 3643 (1999).
\bibitem{slichter}C.P. Slichter, Principles of Magnetic Resonance, (Springer-Verlag, New York 1989),
3rd ed.
\bibitem{shimizu}T. Shimizu et al., J. Phys. Soc. Jpn. {\bf 62}, 772 (1993).
\bibitem{shielding}The two important parameters needed for the simulation are the 
experimentally determined onsite and offsite shielding factors \cite{slichter}.
We self-consistently deduced them
by inputting the temperature dependence of the lattice constants 
from neutron diffraction [M. Braden et. al., Physica C {\bf 223}, 396 
(1994)] into the EFG 
calculation and directly compared it to our 
$^{63}\nu _{Q}$ (peak resonance frequency) data over the same 
temperature range. Using the same notation as \cite{shimizu}, we 
deduce the shielding factors for onsite $(1-\gamma _{R}) <r^{-3}> = 5.40 $ a.u.
and offsite $(1-\gamma_{\infty}) = 18.9$ contributions, both of which are in close agreement
with previous estimates \cite{shimizu} deduced using an entirely different method.
\bibitem{haskel}D. Haskel et al., Phys. Rev. Lett. {\bf 76}, 439 (1996).
\bibitem{bozin}E.S. Bozin et al., Phys. Rev. B {\bf 59}, 4445 (1999).
\bibitem{braden}M. Braden et. al., Phys. Rev. B {\bf 63}, 140510 (2001).
\bibitem{tilt}Tilting the CuO$_{6}$ octahedron away from the c-axis with 
angle $\theta$ results in a shift in the Cu resonance frequency given by
$\bigtriangleup \nu \sim \theta ^{2}$.
\bibitem{takagidelta}H. Takagi et. al., Phys. Rev. Lett. {\bf 68}, 3777 (1992).
\bibitem{yamada}K. Yamada et. al., Phys. Rev. B {\bf 57}, 6165 (1998) 
and references there in.
\bibitem{ino}A. Ino et. al., Phys. Rev. Lett. {\bf 79}, 2101 (1997).
\bibitem{chen}C.Y. Chen et. al., Phys. Rev. B {\bf 43}, 392 (1991). 
\end{references}
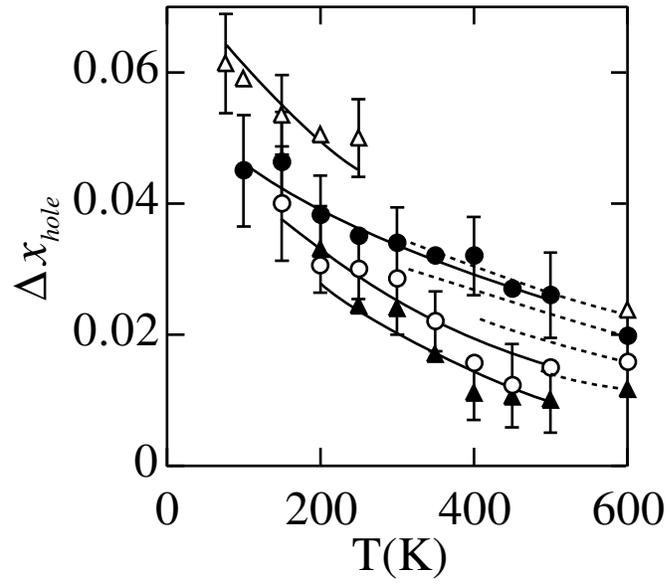
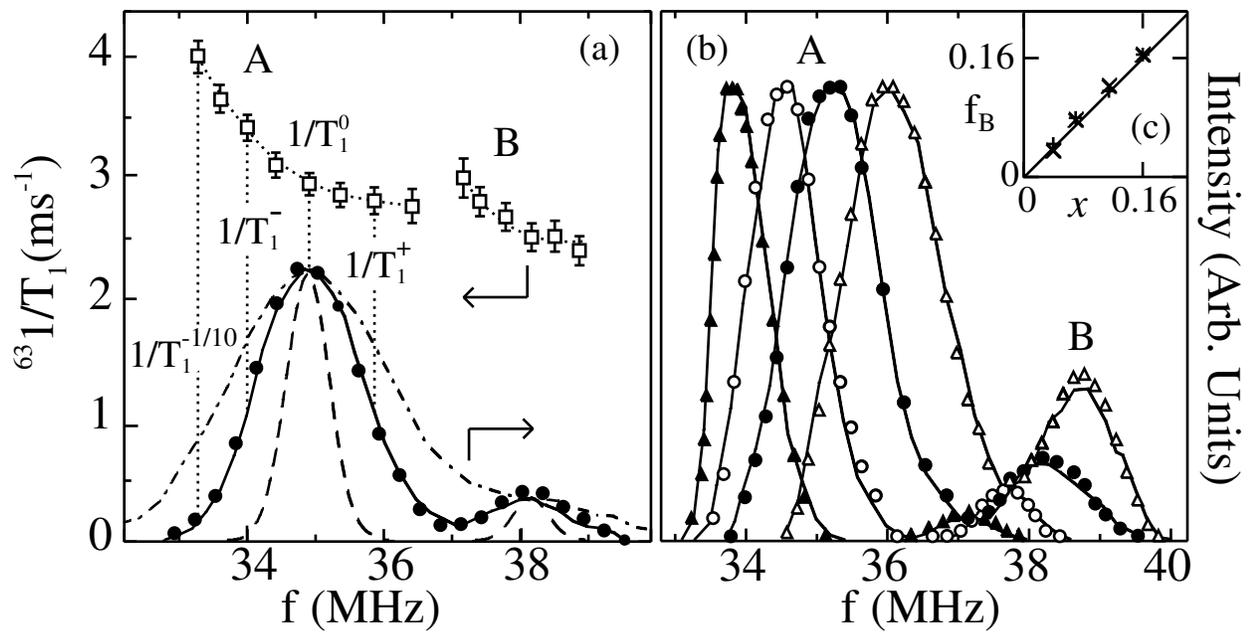
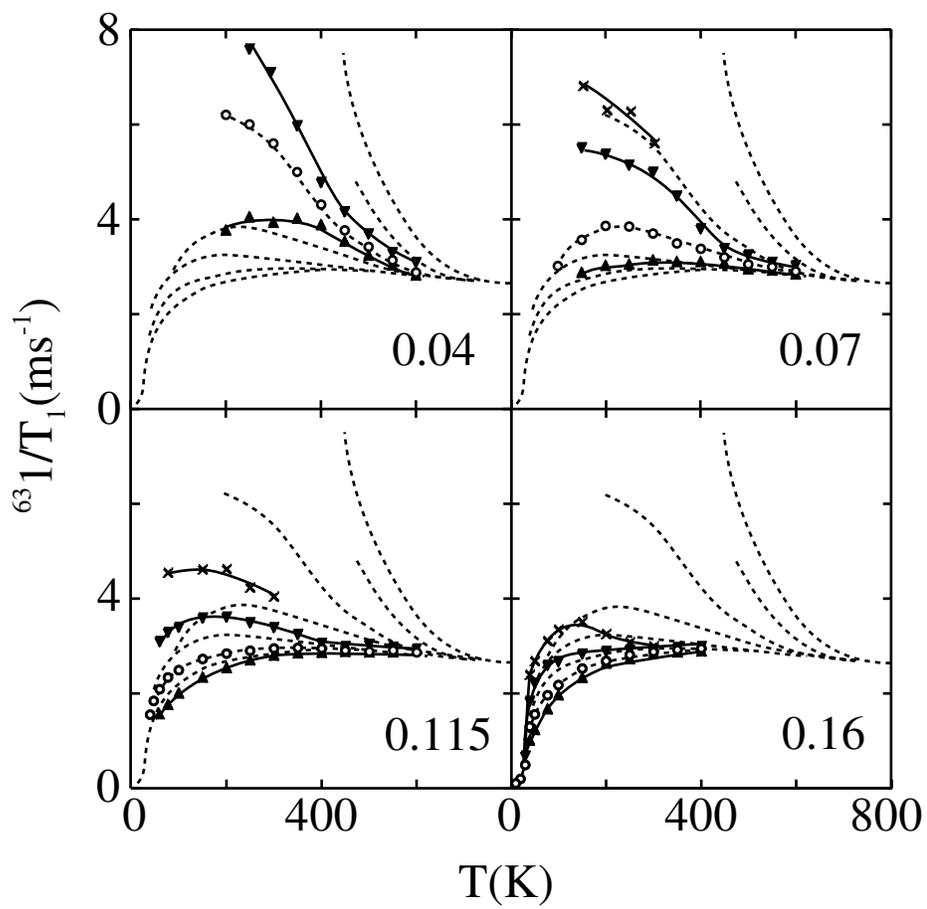
\end{document}